\documentclass[fleqn,10pt]{wlscirep}
\usepackage[utf8]{inputenc}
\usepackage[T1]{fontenc}

\newcommand{\figref}[1]{Fig.\ \ref{#1}}

\title{Directed Cycles as Higher-Order Units of Information Processing in Complex Networks}

\author[1,2,*]{Hardik Rajpal}
\author[3,+]{Paul Expert}
\author[4,5,+]{Vaiva Vasiliauskait\.e}
\affil[1]{Centre for Complexity Science, Imperial College London, London, SW7 2AZ, United Kingdom}
\affil[2]{Department of Mathematics, Imperial College London, London, SW7 2AZ, United Kingdom}
\affil[3]{Global Business School for Health, Faculty of Population Health Sciences, University College London, WC1E 6BT, UK}
\affil[4]{Laboratory of Biosensors and Bioengineering, Institute for Biomedical Engineering, Gloriastrasse 37/39, ETH Z\"urich, Z\"urich, 8092, Switzerland}
\affil[5]{Computational Social Science, Stampfenbachstrasse 48, ETH Z\"urich, Z\"urich, 8092, Switzerland}

\affil[*]{h.rajpal15@imperial.ac.uk}

\affil[+]{these authors contributed equally to this work}

\begin{abstract}
Directed cycles form the fundamental motifs in natural, social and artificial networks, yet their distinct computational roles remain under-explored, particularly in the context of higher-order structure and function. In this work, we investigate how two types of directed cycles — feedforward and feedback — can act as higher-order structures to facilitate the flow and integration of information in sparse random networks, and how these roles depend on the environment of the cycles. 
Using information-theoretic measures, we show that network size, sparsity and relative directionality critically impact the information-processing capacities of directed cycles. In a network with no-preferred global direction, a feedforward cycle enables greater information flow and a feedback cycle allows for increased information integration. The relative direction of a feedforward cycle as well as the structural incoherence it induces, determines its capacity to generate higher-order behavior.
Finally, we demonstrate that introducing feedback loops into otherwise feedforward architectures increases the diversity of network activity patterns. These findings suggest that directed cycles serve as computational motifs with local information processing capabilities that depend on the structure they are embedded.
Using directed cycles, we highlight the interdependence between higher-order structures and the higher-order order behaviour they can induce in the network dynamics.
\end{abstract}

\begin{document}

\flushbottom
\maketitle
%
%
\thispagestyle{empty}

\section{Introduction}

Network science has been instrumental in understanding emergent phenomena across domains as diverse as social systems, 
technological infrastructures, and the human brain. Traditional graph-based models, built on pairwise interactions have provided deep insights into processes such as contagion~\cite{brockmann2013hidden}, synchronization~\cite{arenas2008synchronization,gomez2007paths}, and resilience~\cite{albert2000error}. Recent advances in higher-order network science have introduced mathematical frameworks, such as simplicial complexes~\cite{Petri:2014hq}, hypergraphs~\cite{bianconi2024theory}, and higher-order Markov models~\cite{lambiotte2019networks}, where the fundamental building blocks are not limited to representing pairwise interactions~\cite{BATTISTON20201,bick2023, majhi2022dynamics}. These objects can therefore be used to represent new types of interaction mechanisms to generate emergent behaviours that cannot be fully explained by dyadic interactions alone: phenomena such as group synchronization\cite{millan2020explosive,Arnaudon_2022,Nurisso_2024,Carletti_2023}, collective decision-making~\cite{burgio2020evolution}, and higher-order contagion~\cite{Iacopini_2019,ferraz2024contagion}. 
At the same time, a growing body of work emphasizes the distinction between higher-order structures, or interactions~\cite{Battiston_2021,rosas2022disentangling}, and higher-order behaviors, showing that the latter can also emerge from specific arrangements of pairwise interactions. For example, signed pairwise networks can generate group-level frustration in systems of Ising spins~\cite{rosas2022disentangling}.


In this paper, we explore the relationship between higher-order structures and behaviors further by presenting a mechanism to tune the strength of the higher-order behaviors elicited by a given higher-order structure in the context of directed networks. Directedness is a particularly important property in systems that process information, and the relative organisation of directed edges defines the function of a system. Directed networks have been crucial in guiding our understanding of the informational architecture of the brain~\cite{bassett2017network}, the propagation of social influence~\cite{borgatti2009network}, and the design of AI systems such as feedforward and recurrent neural networks~\cite{mocanu2018scalable}. Directed cycles play a crucial role in modulating the dynamics within these networks. Unlike undirected cycles, different directed cycles can impact the flow of information on the network in fundamentally different ways, acting as minimal higher-order structures that introduce memory, path dependencies, and non-Markovian dynamics, thereby shaping both local information processing and global system behaviour~\cite{vasiliauskaite2022cycle}.

The primary visual regions of the brain are organized in a feedforward manner, where information flows from lower to higher areas of the visual cortex~\cite{markov2014anatomy}. These feedforward cycles create multiple copies of the same information, ensuring robust information transfer through increased redundancy~\cite{luppi2022synergistic}. In contrast, regions of the visual cortex higher in the hierarchy often exhibit a higher degree of recurrent feedback cycles to facilitate complex processes like attention~\cite{markov2014anatomy}. These feedback cycles are thought to integrate information in time and from different parts of the network, allowing for more higher-order synergistic interactions~\cite{luppi2022synergistic}. Beyond the brain, in scale-free social networks, feedback cycles have been shown to play a crucial role in facilitating echo chambers and polarization~\cite{rajpal2019tangled}. Feedforward and recurrent neural network architectures have been shown to be effective in different tasks, indicating that the type of directed cycles present in the network can have a significant impact on the network's computational capabilities~\cite{van2020going}.


Most studies on cyclic network motifs focus on the relative abundance of the different motifs and their relationship to the observed dynamics in the network. Few have focused on identifying the distinct impact of different types of directed cycles on network dynamics. For instance, it has been shown that the presence of feedforward and feedback cycles impacts the synchronizability of a network~\cite{lizier2023analytic}. In directed ring networks, an abundance of feedback cycles relative to feedforward cycles can facilitate synchronization at lower coupling strengths~\cite{brede2008synchronization}. Some studies have also tried to map the computational capabilities of directed cycles: the in-degrees of source and target nodes in a feedforward cycle can modulate the amount of information transfer~\cite{novelli2020deriving}, and structurally balanced directed cycles can improve the performance of sparse recurrent neural networks~\cite{zhang2023universal}.


Building on these studies, we aim to investigate triad higher-order structures, focusing on how their characteristics and their embedding within a network environment shape their functional roles. We focus on two types of directed cycles —feedforward and feedback— and examine how their embedding in a directed network shapes both their local computational roles and their contribution to the network’s global dynamical properties. Using information-theoretic measures, we systematically analyze how the interaction between a directed cycle and its surrounding network determines its dynamics in general, and its ability to generate higher-order behavior in particular. 

We find that feedback and feedforward cycles have intrinsically different capabilities for generating higher-order behavior, but that the magnitude of these capabilities depends strongly on the properties of the environment in which they are embedded. In particular, the relative difference between the local directionality of a cycle and the global directionality of the host network can be tuned to modulate the amount of higher-order behavior produced. Our results indicate that the role a higher-order structure plays in the dynamics of a system is determined not only by the structure itself, but also by how it is embedded in its environment. 
Furthermore, localisation and abundance of feedforward and feedback cycles in sparse networks can be used to control the diversity and nature of a network dynamics. We believe the relationships between structures an be directly translated in both   understanding and designing of biological and artificial systems. 

\section*{Results}

\begin{figure}[!h]
    \centering
    \includegraphics[width=1\linewidth]{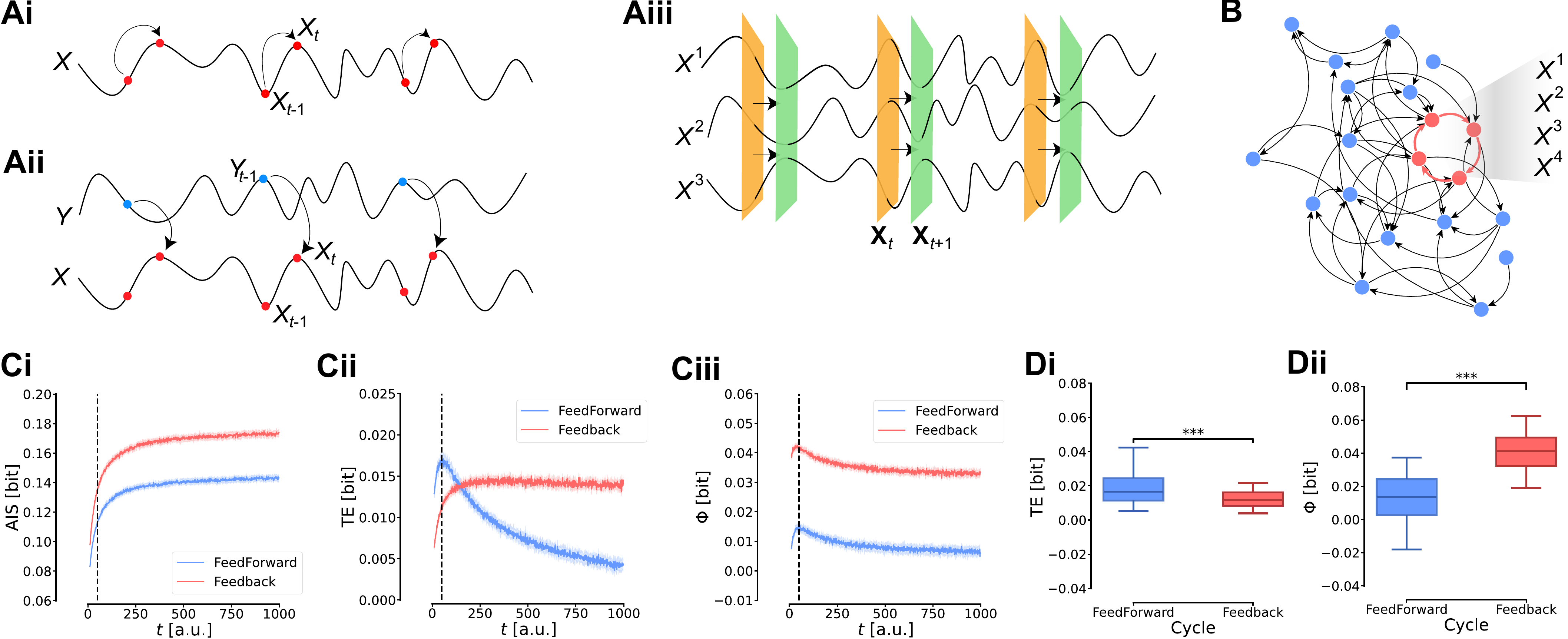}
    \caption{\textbf{Information-theoretic measures and their expression in cycles embedded into directed Erd\"os-R\'enyi graphs.} 
\textbf{A} Schematic representation of the information-theoretic measures used in this study. 
\textbf{Ai} \textit{Active Information Storage} (AIS): predictive information stored in the past of a variable ($X_{t-1}$) about its current state ($X_t$). 
\textbf{Aii} \textit{Transfer Entropy} (TE): information provided by the past of a source ($Y_{t-1}$) about the present of a target ($X_t$), beyond that stored in the target’s own past. 
\textbf{Aiii} \textit{Integrated Information} ($\Phi$): excess information stored in the joint state of variables $\textbf{X}_t=\{X^1_t,X^2_t,X^3_t\}$ about their joint future $\textbf{X}_{t+1}$, beyond the information contained in the individual parts ($X^1, X^2, X^3$). 
\textbf{B} These measures are applied to time series extracted from a four-node embedded cycle in a network. 
\textbf{C} Estimates of each measure at every simulation time step, obtained from ensembles of directed Erdős–Rényi networks endowed with either a feedback or a feedforward cycle. 
AIS grows to a steady state (\textbf{Ci}), TE in the feedforward cycle peaks during the transient phase (\textbf{Cii}), and $\Phi$ peaks during the transient in both cycle types (\textbf{Ciii}). 
\textbf{D} Information-processing characteristics at the transient peak: TE is significantly larger in feedforward cycles, whereas $\Phi$ is larger in feedback cycles.}
    \label{fig:1}
\end{figure}

We quantified spatiotemporal patterns of computation and information processing in networks using information-theoretic measures applied to simulated dynamics. Each network consisted of nodes $i$ associated with (potentially multivariate) time-varying variables $X^i(t)$, connected to a set $\mathcal{N}_i$ of neighbors via directed edges. Network dynamics followed a stationary first-order multivariate vector autoregressive [VAR(1)] process, where each node’s state depends linearly on its own past and that of its neighbors, plus Gaussian noise (see Methods). The coupling strength was set to satisfy the stability condition, ensuring that activity relaxed to a steady state. From the simulated time series, we computed active information storage (AIS) (\figref{fig:1}Ai), which quantifies the predictive influence of a node’s own past, and transfer entropy (TE) (\figref{fig:1}Aii), which captures directed influences from neighboring nodes.

We applied these measures to cycles embedded in the network, treating them as composite computational units. We focus on two directed cycle types—feedforward and feedback—which represent fundamentally different modes of activity propagation in directed networks~\cite{vasiliauskaite2022cycle}. Based on their topology, we expect feedforward cycles to promote robust information transfer through redundancy and parallel paths, whereas feedback cycles should support temporal integration by combining inputs from different network regions while retaining memory of past activity. To capture higher-order collective behavior within these cycles, we compute integrated information $\Phi$ (\figref{fig:1}Aiii), which quantifies the excess information that the cycle has about its joint future beyond what is available from its parts—edges—considered individually.

The computational nature of our framework enables a systematic exploration of how key network parameters shape information processing in feedforward and feedback cycles. We vary network size, density, and a global directionality parameter to assess their effects on both cycle types. By embedding cycles at different positions within the network, we examine how structural context modulates integrated information and transfer entropy. Beyond local processing, we link these dynamics to global network behavior by applying graph signal processing to the simulated activity, quantifying the diversity of network harmonic modes. Together, these analyses allow us to directly compare the computational roles of feedforward and feedback cycles under controlled conditions.

\subsection{Feedforward and feedback cycles exhibit contrasting information-processing capacities}

We first compare the information-processing capabilities of feedforward and feedback cycles embedded in sparse Erdős–Rényi directed networks with no preferred directionality. Each network contains one embedded cycle with either feedforward or feedback directionality. From an ensemble of 10,000 simulated trajectories, we extract the activity of the nodes forming each cycle and compute time-resolved measures of active information storage, transfer entropy, and integrated information~\cite{gomez2015assessing}. To isolate the specific contribution of the cycle, we compare these values to those obtained from a null model in which the cycle is replaced by a disconnected pair of edges (see \ref{sec:rannets} in Methods for description of the null model).

AIS—the total information shared between the past and current state of a cycle—is significantly higher for the feedback cycle than for the feedforward cycle in the steady state (Figure~\ref{fig:1}Ci). Following the integrated information decomposition framework~\cite{mediano2021towards}, AIS can be decomposed into non-overlapping modes of information processing: TE, capturing information transfer, and $\Phi$, capturing information integration. Because the VAR model implements a diffusion-like process, most information transfer occurs during the transient regime, and AIS increases until steady state. The end of this transient period marks the richest and most dynamic phase of information processing; once steady state is reached, information dynamics stabilize. We therefore characterize the information-processing capacity of each cycle by the maximum value of each measure during the transient.

When comparing these temporal maxima, feedforward cycles exhibit higher TE (Figure~\ref{fig:1}Cii), whereas feedback cycles show greater $\Phi$ (Figure~\ref{fig:1}Ciii). Across 1,000 independent network realizations, these differences are statistically significant for both measures (Figure~\ref{fig:1}Di,Dii), confirming that the two cycle types occupy distinct roles in network information processing.

\subsection{Strength of information processing depends on cycle’s positioning and network’s global structure}

\begin{figure}[h!]
    \centering
    \includegraphics[width=1\linewidth]{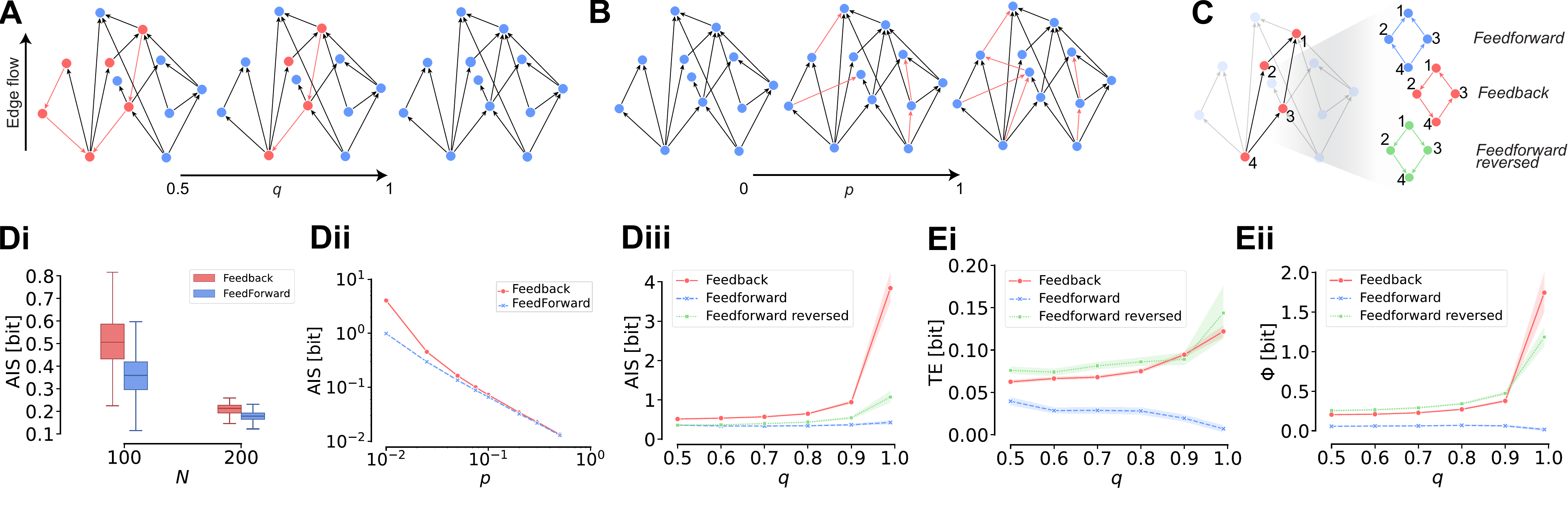}
   \caption{\textbf{Effect of network directionality on the information processing of embedded cycles.} 
We vary directed Erdős–Rényi graphs with a preferred directionality by systematically changing three parameters: 
network size ($N$), density ($p$), and global directionality ($q$). 
\textbf{A} Probability $q$ controls the alignment of edges with the global flow: $q = 1$ corresponds to complete alignment, while $q = 0.5$ indicates no preferred direction. 
\textbf{B} Increasing $p$ adds more edges, producing denser directed networks. 
\textbf{C} Cycles embedded in a network with a preferred direction can be aligned feedforward, embedded as feedback, or aligned feedforward in the reversed direction. 
\textbf{D} AIS decreases with increasing $N$ (\textbf{Di}) and $p$ (\textbf{Dii}), but increases with $q$ for feedback and reversed feedforward cycles (\textbf{Diii}). 
\textbf{E} Decomposing AIS into TE (\textbf{Ei}) and $\Phi$ (\textbf{Eii}) shows that both components increase with $q$ for feedback and reversed feedforward cycles.
}
    \label{fig:2}
\end{figure}

The information-processing capacity of a cycle is shaped not only by its topology but also by the structural properties of the network in which it is embedded. To quantify these effects, we examine how active information storage (AIS) varies with network size ($N$), density ($p$), and relative global directionality ($q$). Across network sizes and densities, AIS decreases as $N$ increases (Figure~\ref{fig:2}Di), reflecting the fact that a cycle captures a larger fraction of the diffusion process in smaller networks. AIS also decreases with increasing $p$ (Figure~\ref{fig:2}Dii), consistent with the cycle exerting a stronger influence on information flow in sparser networks.

Finally, to understand how the global directionality of the network affects the information processing capabilities of the cycles, we systematically vary the 
the flow of edges on a random directed graph. The direction of each edge is assigned with probability $q$ from the lower- to the higher-indexed node, such that $q = 1$ yields a fully acyclic network and $q = 0.5$ corresponds to no preferred direction. In this setting, a feedforward cycle can be embedded either aligned with the global flow or in the reversed orientation, alongside the feedback cycle. 

As $q$ increases, AIS rises for both the reversed feedforward and feedback cycles, but remains comparatively unchanged for the aligned feedforward cycle (Figure~\ref{fig:2}Diii). Decomposing AIS shows that $\Phi$ and TE increase with $q$ for the reversed feedforward and feedback cycles (Figure~\ref{fig:2}Ei,Eii), whereas the aligned feedforward cycle exhibits negligible values of $\Phi$ and a decline in TE. These results indicate that a cycle’s capacity to process information and act as a higher-order structure depends critically on its ability to disrupt and redirect the prevailing flow of information in the network.

\subsection{Cycle-induced perturbations to trophic coherence shape information processing}

To assess how embedded directed cycles disrupt the directional organization of a network, we quantify trophic coherence~\cite{mackay2020directed}. This measure is based on assigning each node a trophic level and counting the fraction of edges that violate the trophic ordering, defined as the global incoherence $F_0$ ($0$ = fully coherent, $1$ = maximally incoherent, see Methods).

\begin{figure}[h!]
    \centering
    \includegraphics[width=1\linewidth]{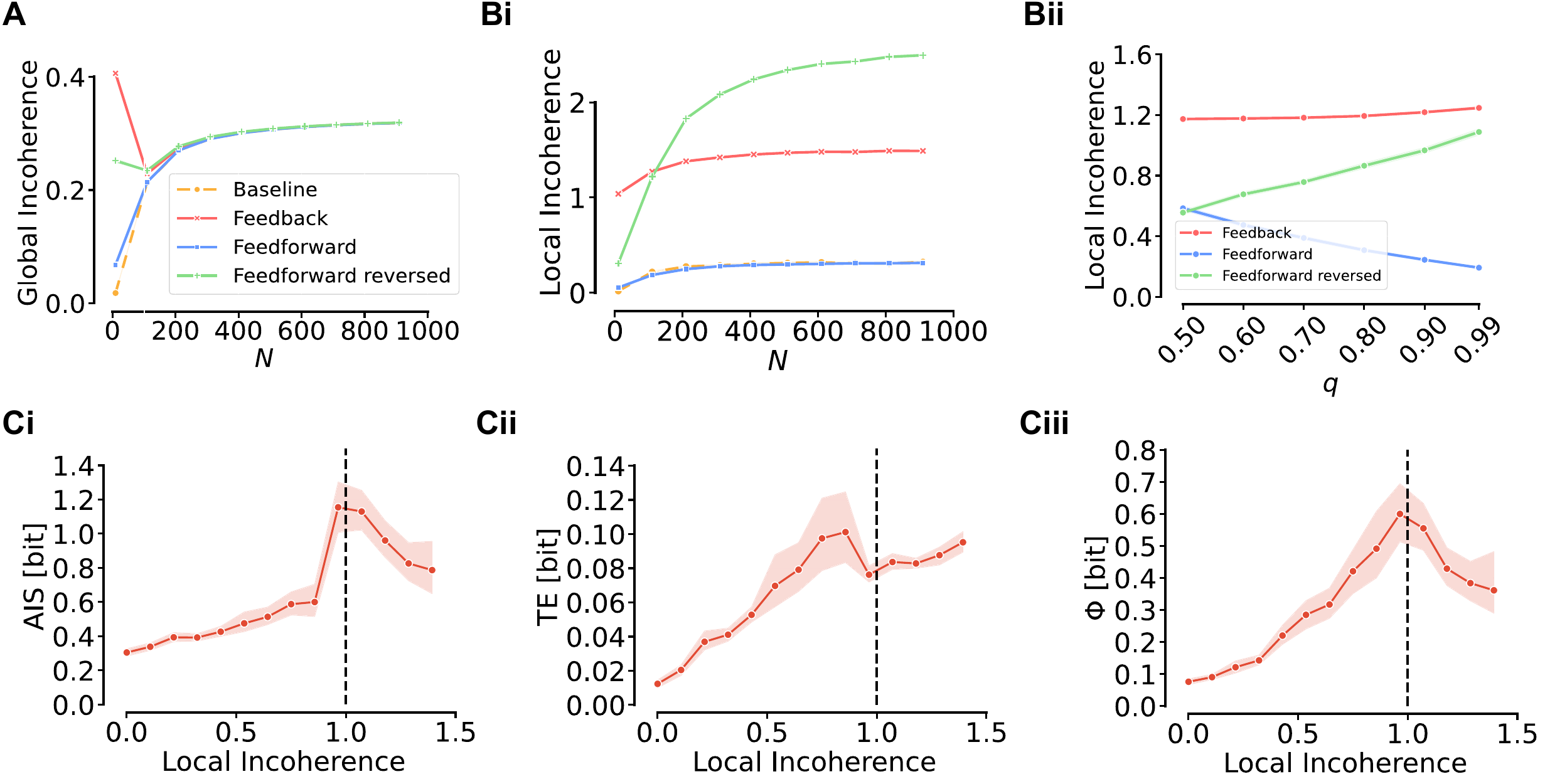}
    \caption{\textbf{Cycle-induced perturbations to trophic coherence and their impact on information processing.} 
    \textbf{A} Global trophic incoherence ($F_0$) for networks of varying size $N$ with embedded feedback, feedforward, or reversed feedforward cycles, compared to baseline networks. Small networks show a marked increase in $F_0$ when a cycle is added, but this effect diminishes with size. 
    \textbf{Bi} Local incoherence ($F_C$) for the same networks, computed using only the edges within the cycle. For large networks, $F_C$ is highest for reversed feedforward cycles, followed by feedback cycles, with aligned feedforward cycles showing baseline-level $F_C$. 
    \textbf{Bii} Dependence of $F_C$ on global directionality $q$: reversed feedforward cycles increase linearly with $q$, feedback cycles increase sublinearly, and aligned feedforward cycles decrease linearly. 
    \textbf{Ci–Ciii} Relationship between $F_C$ and (Ci) active information storage (AIS), (Cii) transfer entropy (TE), and (Ciii) integrated information ($\Phi$). AIS and $\Phi$ peak near $F_C \approx 1.0$, while TE peaks at slightly lower $F_C$.}
    \label{fig:3}
\end{figure}

For small networks, adding a single directed cycle markedly increases $F_0$ (Figure~\ref{fig:3}A), but the effect diminishes as system size grows. To isolate cycle-specific effects independent of network size, we define a local incoherence $F_C$ by counting only edges within the cycle that violate the trophic ordering. Here, $F_C = 0$ means that the cycle is perfectly aligned with the network’s trophic hierarchy, whereas $F_C \approx 1$ means that, on average, each edge in the cycle violates the ordering once. Values $F_C > 1$ indicate that multiple violations occur per edge when the cycle is considered within the network’s trophic framework, corresponding to a strong local disruption of directional flow.

Across large networks, $F_C$ is highest for the reversed feedforward cycle, followed by the feedback cycle, while the aligned feedforward cycle shows no deviation from the baseline (Figure~\ref{fig:3}Bi). Varying the global directionality $q$ reveals that $F_C$ increases linearly for the reversed feedforward cycle, sublinearly for the feedback cycle, and decreases linearly for the aligned feedforward cycle (Figure~\ref{fig:3}Bii).

Finally, relating $F_C$ to the information-theoretic measures shows that AIS and $\Phi$ both increase with $F_C$ and reach their maximum at $F_C \approx 1.0$, whereas TE peaks at a slightly lower $F_C$ (Figure~\ref{fig:3}Ci–Ciii). This indicates that a cycle’s capacity for information storage and integration is greatest when its edges collectively introduce, on average, one trophic violation—striking a balance between forward and backward flows. In contrast, information transfer is optimal under a slightly more ordered configuration, short of being fully acyclic or tree-like. The non-monotonic trends further suggest that excessive disruption to the directional flow, approaching random connectivity, can be detrimental to all forms of information processing.


\subsection{Local cycle structure modulates the diversity of global network dynamics}

\begin{figure}[!h]
    \centering
    \includegraphics[width=1\linewidth]{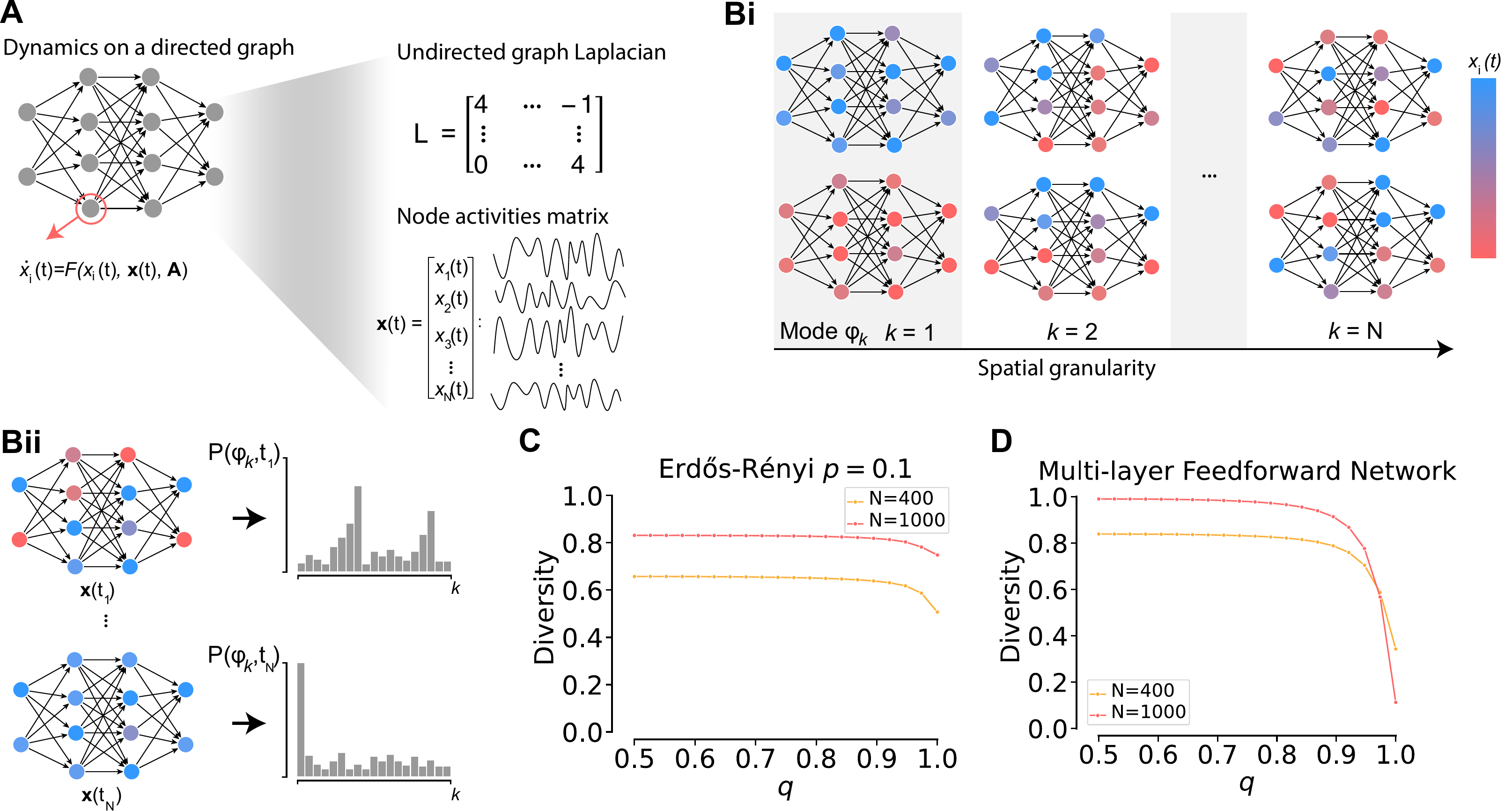}
    \caption{
\textbf{Directed cycles enhance the diversity of global network dynamics.} 
\textbf{(A)} Network harmonic decomposition combines the eigenvectors of the symmetrized graph Laplacian with node activity time series from the VAR dynamics on a directed graph, yielding spatial modes of network activity. 
\textbf{(Bi)} Modes are ordered by eigenvalue $k$: low-$k$ modes correspond to global activity, while high-$k$ modes capture localized fluctuations. 
\textbf{(Bii)} The power spectrum across modes reveals how strongly each mode contributes to the network’s dynamics at any time; the Shannon entropy of this distribution quantifies diversity.  
\textbf{(C)} In directed Erdős–Rényi graphs, diversity increases as the network becomes less directional ($q \to 0.5$) and more feedback cycles appear; the effect is stronger in larger networks.  
\textbf{(D)} Multi-layer feedforward networks show a similar trend, with even greater gains in diversity as directionality decreases.
}
    \label{fig:4}
\end{figure}

Beyond their role in local information processing, directed cycles influence the repertoire of activity patterns a network can sustain. We quantify this global impact using \emph{network harmonic decomposition}~\cite{atasoy2018harmonic}, a Graph Signal Processing (GSP) approach that expresses the network’s activity as a sum of spatial modes (Figure\ref{fig:4}A,B). Lower-index modes capture broad, system-wide fluctuations, whereas higher-index modes reflect localized patterns restricted to smaller subsets of nodes.
The relative power across these modes provides a measure of the diversity of global activity patterns—higher diversity indicates that activity is distributed across many spatial scales rather than dominated by a few. Such diversity has been linked to functional richness in biological networks, including correlations with levels of consciousness~\cite{luppi2023distributed} and the organization of resting-state brain networks~\cite{atasoy2016human}, as well as to the phases of the XY model~\cite{Expert:2017cz}.
Further details of the network harmonic decomposition can be found in the Methods section~\ref{sec:gsp}.

We assess how directed cycles modulate this diversity by embedding them into networks with systematically varied global directionality $q$—from fully aligned flows ($q=1$) to no preferred direction ($q=0.5$)—and comparing two topologies: Erdős–Rényi graphs and multi-layer feedforward networks.
In both cases, diversity increases as $q$ decreases, reflecting the greater diversity of the signal with larger prevalence of cycles opposing the dominant flow (Figures~\ref{fig:4}C,D). The effect is amplified in larger networks and particularly pronounced in multi-layer feedforward architectures, where reductions in directionality yield disproportionately large gains in diversity.

Together, these results reveal that local disruptions to directional flow—such as feedback cycles—can substantially enrich the spectrum of spatial activity patterns available to a network, providing a mechanism to enhance its global dynamical repertoire.

\section*{Discussion}
In this work, we investigate the role of directed cycles in local information processing, how the network structure influences their dynamical properties, and their impact on the global activity patterns of the network. We propose local trophic coherence as a measure that explains the relationship between relative directionality of a cycle and the network and how it is a parameter that determines the level of higher-order behavior feedforward cycles can generate. These findings highlight the role of directed feedforward and feedback cycles as higher-order structures which impact both local and global network dynamics.

In this work we focus on directed cycles of length 4. However, other arrangements of directed edges can create other higher-order motifs~\cite{vasiliauskaite2022cycle} with distinct computational properties of any lengths. The measures and the tools presented in this paper can be extended to characterize the role of other higher-order structures in networks. On the other hand, in real-world systems where the network structure is unobserved but the activity of the nodes can be recorded, the information processing signatures can be used to infer the local network structure.

In order to assess the global impact of directed cycles, we adapted the framework of network harmonic decmposition, originally proposed for undirected networks. The framework when applied to time series activity generated from directed graphs, is able to explain the spatial decomposition of the global activity, however at the cost of directionality. While frameworks have been developped~\cite{marques_2020}, general interpretability and applicability is still lacking.

This works confirms the clear distinction that needs to be made between higher-order structure and function~\cite{rosas2022disentangling}, but also highlights that simple alteration to the structure, in our case the relative directionality of a cycle can provide explicit mechanisms to tune high-order behaviour. And while a better understanding of the interdependence between structure and function is necessary to further our understanding of higher-order behaviours in networked systems, we believe this work opens new avenues of research in that direction.
 
In summary, the findings of the paper reveal that directed cycles and other higher-order motifs can be used as computational primitives to embed higher-order interactions in a network, which can significantly enhance the information processing capabilities of the network. This work provides a foundation for future investigations into the role of directed cycles in complex networks, and how they can be used to design more efficient and robust information processing systems.

\section{Methods}

Our analysis proceeds in four main stages:
(i) generating directed network topologies with specified structural properties;  
(ii) embedding a tagged set of nodes into a directed cycle (feedforward, reversed feedforward, or feedback);  
(iii) simulating network activity using a stationary first-order vector autoregressive (VAR(1)) process; and  
(iv) quantifying both local and global information-processing properties using information-theoretic and graph-spectral measures.

For each simulation, we generated a network with prescribed size, density, and global directionality. A selected cycle was embedded by tagging four nodes and connecting them in a specific directed configuration. Network dynamics were simulated under the VAR(1) process from random initial conditions, producing ensembles of 10{,}000 trajectories per network. From these, we extracted the activity of the cycle nodes, along with the full network time series when required. The cycle trajectories were used to compute time-resolved measures of active information storage (AIS), transfer entropy (TE), and integrated information ($\Phi$), and compared to a null model in which the cycle was replaced by a disconnected pair of edges. In parallel, we computed network-level metrics, including global and local trophic coherence, and performed network harmonic decomposition to assess the diversity of spatial activity patterns. Each of these steps is described in detail below, in the order required to reproduce the analysis.

\subsection{Network models}
\label{sec:rannets}

We considered two directed random network models, each parameterized by size ($N$), density ($p$), and global directionality ($q$), along with a null model.

\paragraph{Directed Erd\"os–R\'enyi model.}
Starting from the classic Erd\"os–R\'enyi random graph $G(N,p)$, we assigned a unique integer index to each node and connected each unordered pair with probability $p$. For each existing edge $(u,v)$, direction was assigned from the lower to the higher index with probability $q$, and reversed otherwise. When $q = 1$, the graph is a directed acyclic graph with a perfectly ordered flow; when $q = 0.5$, edge directions are random with no preferred global orientation.

\paragraph{Directed multi-layer feedforward model.}
Nodes were arranged in $L$ layers, with edges between successive layers assigned independently with probability $p$. Directionality was imposed by orienting each feedforward edge from the earlier to the later layer with probability $q$, and flipping its direction otherwise. Thus, $q = 1$ yields a fully feedforward architecture, whereas $q = 0.5$ produces no preferred flow between layers.

\paragraph{Null model.}
For information-theoretic analyses, we constructed a null model in which the embedded cycle was replaced by a pair of disconnected edges with the same source and target. This preserves local degree and global topology, but removes the higher-order cyclic structure.

\subsection{Directionality of a graph}
\label{sec:graphdir}

In a directed network, global directionality reflects the extent to which edges align with a consistent flow without forming feedback loops.  
A fully ordered directed acyclic graph (DAG) is maximally coherent, while a randomly oriented network is less coherent.

We quantify this property using the trophic coherence framework~\cite{mackay2020directed}.  
First, each node $n$ is assigned a trophic level $h_n$ by solving
\begin{equation}
    \Lambda h = v,
\end{equation}
where $v_n = \sum_m A_{mn} - \sum_m A_{nm}$ and  
$\Lambda = \mathrm{diag}(u) - A - A^{\top}$ with $u_n = \sum_m (A_{mn} + A_{nm})$.  
The global trophic incoherence is
\begin{equation}
    F_0 = \frac{\sum_{mn} A_{mn} (h_n - h_m - 1)^2}{\sum_{mn}A_{mn}},
\end{equation}
ranging from $F_0=0$ (maximally coherent) to $F_0=1$ (maximally incoherent).  

\paragraph{Local trophic coherence of a cycle}

To quantify the extent to which an individual cycle disrupts global directionality, we define the local trophic incoherence
\begin{equation}
    F_C = \frac{\sum_{mn \in C} A_{mn} (h_n - h_m - 1)^2}{\sum_{mn \in C}A_{mn}},
\end{equation}
where $C$ is the set of edges in the cycle.  
$F_C = 0$ indicates perfect alignment with the global trophic ordering; higher values correspond to greater local disruption.  
Unlike $F_0$, $F_C$ can exceed $1$ because it is computed over a restricted subset of edges, allowing for stronger violations within the cycle than in the network as a whole.

\subsection{Network dynamics}
Multivariate vector autoregressive models (MVAR) have been extensively used in physical sciences and econometrics to quantify the interdependencies 
in a time-evolving multivariate system. Analytical traceability and efficient computational implementation make the MVAR model a powerful tool for 
exploring multivariate linear dependencies in a complex system.

For our analysis we consider a stationary first-order vector autoregressive (VAR (1)) process defined as 
\begin{equation}
    \textbf{x}_{t+1} = \alpha \textbf{x}_t\textbf{A}+\tau {\varepsilon}_{t} , \quad \varepsilon_{t,i} \sim \mathcal{N}(0,1)
\end{equation}
with the stability condition,
\begin{equation}
    \alpha= \frac{1}{\lambda_{\text{max}}\mathbf{A}}
\end{equation}
Here, $\mathbf{A}$ represents the directed adjacency matrix of the network on which the dynamics are being simulated. For each simulation, 
we extract the trajectories of the four tagged nodes that form the feedback/feedforward cycle for the information-theoretic analysis.

\subsection{Information-theoretic measures}
\label{sec:infomeasures}
Once we have the trajectories of tagged nodes from the cycles, we can analyze these to quantify various temporal information processing measures. 
For our analysis, we focus on three complementary metrics: \textit{Information Storage}, \textit{Integrated Information} and \textit{Transfer Entropy}. 
These metrics are all based on calculating mutual information in multivariate systems. We briefly define these metrics below for a system $X_t$, where $I(X; Y)$ 
refers to the mutual information between $X$ and $Y$.

\textbf{Active Information Storage} (AIS), quantifies the total amount of mutual information between the past and the future of the system. This information can 
further be decomposed into information integration, transfer and other information-specific metrics using the integrated information decomposition~\cite{mediano2021towards}. 
AIS can be simply written for a multivariate system $X$ evolving in time $t$ as,

\begin{equation}
    AIS(X_t) = I(X_t; X_{t+1})
\end{equation}

\textbf{Integrated information} ($\Phi$), over the years various measures of integrated information have been proposed to quantify the amount of excess information 
that a system possesses about its joint future activity as compared to its individual parts~\cite{balduzzi2008integrated,mediano2018measuring}. More recently,
practical measures of integrated information have been proposed that can be computed in a computationally efficient manner on time-series data~\cite{barrett2011practical}.
We use the integrated information measure proposed by Barrett and Seth~\cite{barrett2011practical} $\Phi_E$ which measures the extra information that a system $X$ possesses 
about its joint future activity, compared to the information that can be explained by the minimum information bipartion $\mathcal{B} = {M^1,M^2}$ of the system. 
This measure can be written as, 

\begin{equation}
    Phi_{E}(X_t,\mathcal{B}) = I(X_t; X_{t+1}) - \sum_{i=1}^2 I(M_t^i; M_{t+1}^i)
\end{equation}
where $M^i$ are the minimum information bipartitions of the system $X$ at time $t$. The minimum information bipartition is selected by choosing the partition that minimizes
the normalized integrated information such that,

\begin{equation}
    \mathcal{B} = \arg\min_{\mathcal{B}} \frac{\Phi_E(X_t,\mathcal{B})}{\min_{i} H(M_t^i)}
\end{equation}

However, in systems with a lot of redundancy, there is a lot of shared information among the partitions. 
Thus, the $\sum_{i} I(M_t^i; M_{t+1}^i)$ overcounts this redundant information. This overcounting often leads to a negative value of $\Phi_{E}$ 
for systems with a lot of redundancy. Using integrated information decomposition~\cite{mediano2021towards} a refined measure of integrated information 
is proposed where this redundancy is added back~\cite{luppi2021like}, providing a better scaled measure of integrated information $\Phi_R$.
Here, $\Phi_R$ is defined as with respect to the redundancy function $Red(X_t;X_{t+1})$,

\begin{equation}
    \Phi_R = \Phi_E + (N-1)*Red(X_t;X_{t+1})
\end{equation}

We use the minimum mutual information (MMI) redundancy function to estimate the redundant information among the parts. The MMI redundancy function has been shown to efficiently 
capture the redundancy in multivariate Gaussian systems~\cite{barrett2015exploration}. Here, the MMI redundancy function is defined as,
\begin{equation}
    Red(X_t;X_{t+1}) = \min_{i} I(M_t^i; X_{t+1})
\end{equation}

\paragraph{Transfer Entropy (TE).}  
A conditional mutual information measure quantifying directional influence from a source to a target~\cite{schreiber2000measuring,bossomaier2016transfer}.  
For a VAR(1) process, the past is the current timestep, and the delay $\delta$ is chosen to maximize $TE$:
\begin{equation}
    TE(X^i \rightarrow X^j) = I(X_t^i; X_{t+\delta}^j \,|\, X_t^j).
\end{equation}
In our four-node cycles, source–target pairs are two edges apart, so information propagates in two timesteps; accordingly, we set $\delta=2$, which matches the known peak of $TE$ at the true coupling delay~\cite{wibral2013measuring}.

\subsection{Network Harmonic Decomposition}
\label{sec:gsp}
Fourier transform has enabled the decomposition of a time series in various frequency modes to characterize the activity at different timescales. Similarly, Graph Signal Processing (GSP) 
has enabled Network Harmonic decomposition, which enables the characterization of dynamics at different scales of the network, from local (node level) to global (whole network) and all levels of organization in between. For full reviews and introductions to the field, see \cite{Shuman:2013et, stan2020}
This Network Harmonic Decomposition has enabled the understanding of structure-function correspondence in network dynamics, especially in neuroscience, where it has uncovered scale-specific 
effects in relation to different levels of consciousness~\cite{luppi2023distributed,atasoy2017connectome}.

Here, we use the network harmonic decomposition to decompose the power of the signal into different harmonic modes on the network. Then we use Shannon entropy to calculate the diversity 
of different harmonic modes active on the network. This diversity metric provides a measure of how many different scales of network organization are implicated in the overall dynamics. 
The diversity is high when the power is distributed across a lot of different scales of organization and low when only a few modes are active. The power of each network harmonic mode is 
calculated as described in the previous works~\cite{luppi2023distributed,atasoy2017connectome}, using the eigendecomposition. 

First, the normalized Laplacian matrix ($\tilde{L}$) and its eigenvectors are calculated for the network with adjacency matrix $\mathbf{A}$. The out-degree diagonal matrix of $\mathbf{A}$ 
is defined as $\mathbf{D} = Diag (\sum_j A_{i,j})$, where $\mathbf{D}$ is a matrix (N x N), with diagonal values as out-degrees of the nodes and the off-diagonal elements are zero.

\begin{equation}
    \begin{split}
        &\tilde{L} = D^{-1/2} L D^{1/2} \; \textrm{where} \; L = D - A \\
        &\tilde{L} \varphi_k= \lambda_k \varphi_k
    \end{split}
\label{eq:lap_eig}
\end{equation}
After solving the eigenvalue problem to extract the eigenvector matrix $\varphi_k$, it is convolved with the whole network activity $Y_t$ to estimate the contributions of each 
independent eigenmode ($\omega_k$) using a dot product.

\begin{equation}
    \begin{split}
        &Y_t = \omega_1(t) \varphi_1 + \omega_2(t) \varphi_2 + \dots + \omega_N(t) \varphi_N = \sum_{k = 1}^N \omega_k(t) \varphi_k \\
        &\omega_k(t) = \langle Y_t, \varphi_k\rangle
    \end{split}
\end{equation}

The magnitude or power of the contribution of each harmonic mode $k$ can be calculated from $\omega_k (t)$ at each timestep as,

\begin{equation}
    P_k(t) = \mid \omega_k (t)\mid
\end{equation}

At each timestep, the distribution of power over the different harmonic modes is normalized, $\tilde{P}_k(t) = \frac{P_k(t)}{\sum_{k=1}^N P_k(t)}$, and the Shannon entropy of 
the normalized distribution is calculated as a measure of the harmonic diversity of the dynamics. The measure of diversity $H$ used here is written as

\begin{equation}
    H(t) = - \sum_{k=1}^N \tilde{P}_k(t) \log \tilde{P}_{k}(t) 
\end{equation}
The measure of diversity is calculated at each timestep, and the results present an average value during the steady state. Since the harmonic decomposition requires the
estimation of real eigenvalues, we use the normalized Laplacian matrix $\tilde{L}$ of the symmetrized adjacency matrix of the directed network. However, the activity of 
the nodes is generated using the directed adjacency matrix $\mathbf{A}$, which is used to simulate the VAR dynamics.

\section*{Acknowledgements}
HR is supported by the Statistical Physics of Cognition project funded by the EPSRC (Grant No. EP/W024020/1).
VV acknowledges financial support from the Swiss National Science Foundation and from the European Union Horizon 2020 program (“INFRAIA-01-2018-2019 – Integrating Activities for Advanced Communities,” Grant Agreement No. 871042, SoBigData++: European Integrated Infrastructure for Social Mining and Big Data Analytics, http://www.sobigdata.eu).

\section*{Author contributions statement}
All authors conceived and designed the study. HR and VV developed and implemented the network models. HR performed simulations, and analysed data. PE provided code for graph signal processing. HR and VV generated figures. All authors analysed and interpreted the results, wrote and reviewed the manuscript.

\section*{Competing interests}
Authors declare no competing interests.

\section*{Generative AI statement}
Generative AI tools were used solely for language editing, and improving clarity of expression. All scientific content, data analysis, figures, and interpretations were conceived, developed, and validated entirely by the authors.

\appendix
\renewcommand\thefigure{\thesection.\arabic{figure}}
\clearpage
\section*{Supplementary Information}
\section{Ensemble Mutual Information Calculation}
\setcounter{figure}{0}
In systems where the activity of the parts changes over time, the estimation of the information-theoretic measures cannot be done on a single realisation, as the probability distribution of the activity of the parts changes over time. This problem is well studied in neuroscience experiments where the neural response to a stimulus is often time-varying. In order to assess the coupling dynamics in such systems, an ensemble approach is applied~\cite{gomez2015assessing}. In an experimental study, multiple trials of the same experiment, time-aligned using stimulus timestamp, are treated as an ensemble representing the different realisations of the system~\cite{ince2017statistical}. Similarly, in computational models, multiple realisations of the activity of parts/nodes of the networks can be generated by simulating the model with fixed parameters with different initial conditions. 
Given a sufficient number of realisations, the probability distribution of activity at each time-step yields a stationary probability distribution; these stable probability distributions can then be used to estimate the information-theoretic measures discussed in the paper.

\begin{figure}[h!]
    \centering
    \includegraphics[width=0.8\linewidth]{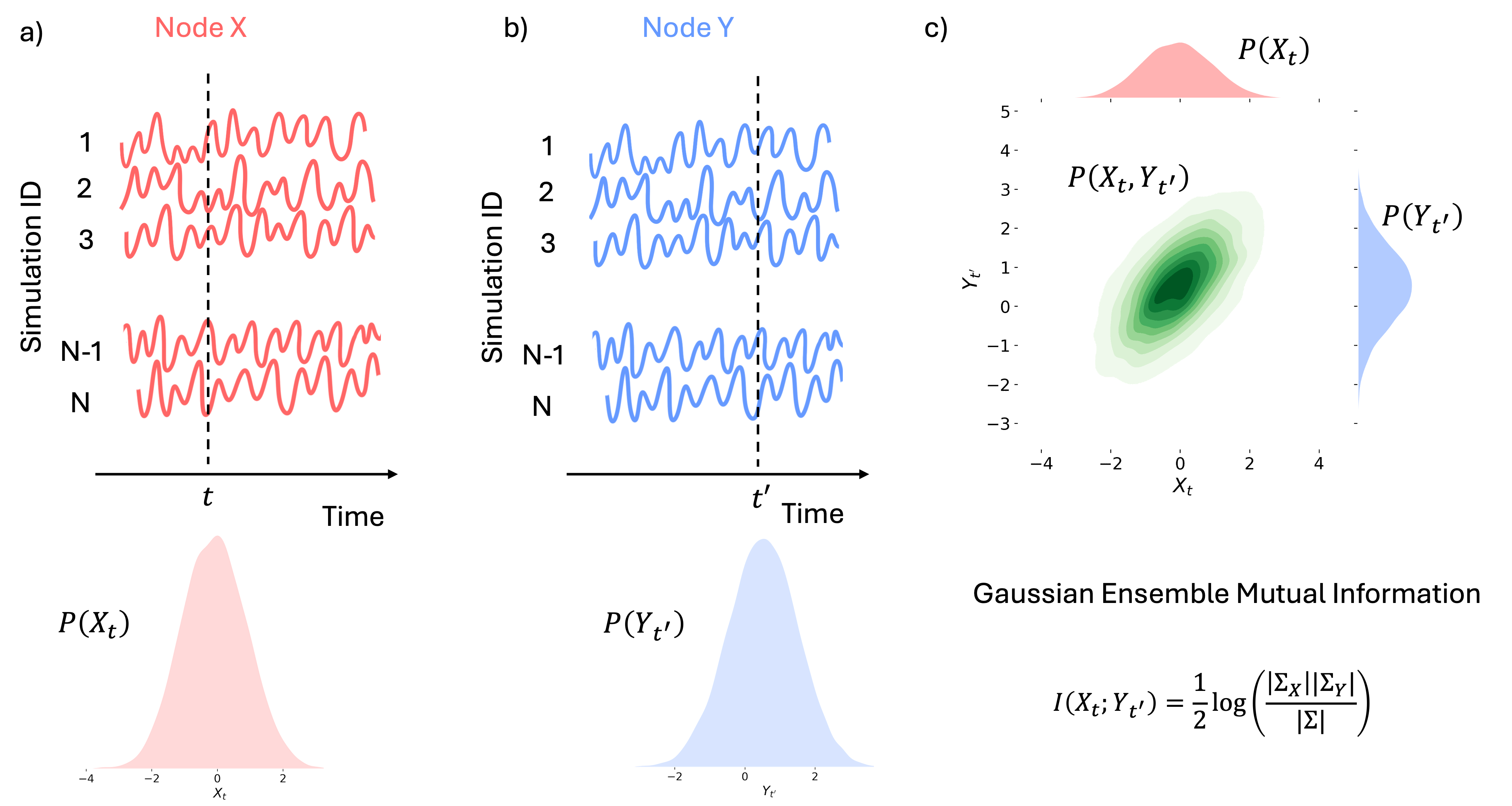}
    \caption[Schematic representation of Ensemble Mutual Information approach]{(a), (b) We track the activity of two nodes $X$ and $Y$ on a network simulated with the same model parameters but with different initial conditions for N different simulations. In order to compute mutual information between the activity of node $X$ at time $t$ and $Y$ at time $t'$, we extract their marginal and joint distributions across the ensemble.(c) Finally we use the joint and marginal distributions to compute mutual information. Under the assumption of Gaussian joint distribution of the variables, mutual information can be computed as a function of their joint ($\Sigma$) and marginal covariances ($\Sigma_X$, $\Sigma_Y$).}
    \label{fig:placeholder}
\end{figure}

\end{document}